\documentclass[aps,prb,reprint,superscriptaddress,showpacs]{revtex4-2}
\usepackage[dvipdfmx]{graphicx}
\usepackage{amsmath}
\usepackage{float}
\usepackage{subfigure}
\usepackage{tikz}
\usepackage[english]{babel}
\usepackage{amssymb}
\usepackage{mathrsfs}
\usepackage{physics}

\usepackage[normalem]{ulem}
\usepackage{xcolor}

\newcommand{\non}{\nonumber}

\def\HFI{H_{\mathrm{FI}}}
\def\Hex{H_{\mathrm{ex}}}

\begin{document}

\title{Spin Seebeck Effect in Graphene}

\author{Xin Hu }
\affiliation{%
Kavli Institute for Theoretical Sciences, University of Chinese Academy of Sciences, Beijing, 100190, China.
}%

\author{Yuya Ominato}
\affiliation{%
Kavli Institute for Theoretical Sciences, University of Chinese Academy of Sciences, Beijing, 100190, China.
}%
\affiliation{%
Waseda Institute for Adacnced Study, Waseda, University, Shinjuku, Tokyo 169-8050, Japan.
}%

\author{Mamoru Matsuo }
\email{mamoru@ucas.ac.cn}
\affiliation{%
Kavli Institute for Theoretical Sciences, University of Chinese Academy of Sciences, Beijing, 100190, China.
}%
\affiliation{%
CAS Center for Excellence in Topological Quantum Computation, University of Chinese Academy of Sciences, Beijing 100190, China
}%
\affiliation{%
Advanced Science Research Center, Japan Atomic Energy Agency, Tokai, 319-1195, Japan
}%
\affiliation{%
RIKEN Center for Emergent Matter Science (CEMS), Wako, Saitama 351-0198, Japan
}%

\date{\today}
\begin{abstract}
We develop a microscopic theory of the spin Seebeck effect (SSE)
at the interface of a bilayer system of a ferromagnetic insulator and graphene.
We compare the tunneling spin current at the interface because of the SSE and the spin pumping (SP), where the SSE and SP are induced by the temperature gradient and the microwave irradiation, respectively.
We demonstrate that the thermally driven SSE exhibits a quantum oscillation pattern similar to that predicted in coherently driven SP. 
Additionally, we show a peak shift of the quantum oscillation owing to the contribution of thermally excited magnons with higher frequencies, which becomes particularly pronounced at higher temperatures.
\end{abstract}

\maketitle

\section{Introduction}
In recent years, the spin Seebeck effect (SSE)\cite{Uchida2008-we,xiao2010,adachi2011} has emerged as a widely recognized phenomenon that converts thermal energy into spin current in various ferromagnets such as a ferromagnetic metal~\cite{Uchida2008-we}, ferromagnetic insulator~\cite{Uchida2010-hr}, ferromagnetic semiconductors~\cite{jaworski2010observation}, and other magnetic materials~\cite{Kikkawa2023-va}.
The discovery of the SSE has led to the establishment of an emergent field, spin caloritronics~\cite{bauer2012spin}, which explores thermoelectric phenomena mediated by electron spins. 
A typical example of SSE is observed in metal/magnet bilayer films~\cite{Uchida2008-we,uchida2010spin}. When a temperature gradient is applied to these films, the magnetic dynamics in the magnetic material are thermally excited. This excitation, through the magnetic interface, transfers angular momentum to the conduction electron spins in the metal, generating a spin current driven by these electrons. 
The generated spin current reflects the spin properties of materials adjacent to the magnet. Accordingly, the SSE can be utilized as a sensitive spin probe for thin films~\cite{Kikkawa2023-va}. 

Spin pumping (SP), a phenomenon closely related to SSE, involves exciting the magnetic dynamics of a metal/magnet bilayer film through microwave irradiation, leading to coherent excitation by ferromagnetic resonance and driving the spins of conduction electrons in the metal via the magnetic interface \cite{mizukami2001study,mizukami2001ferromagnetic,mizukami2002effect,tserkovnyak2002enhanced,vzutic2004spintronics,tserkovnyak2005nonlocal,Hellman2017-fm}. Known for its versatility in injecting spin currents into various materials, SP also provides valuable information about the dynamic spin susceptibility of the material attached to the magnet, serving as a highly sensitive probe for investigating spin characteristics of thin films \cite{han2020spin,qiu2016spin,ominato2020quantum,ominato2020valley,yamamoto2021spin,yama2023effect}, thereby complementing traditional methods like NMR \cite{leggett1975theoretical} and polarized neutron scattering \cite{shull1963Neutron,shull1964Electronic,shull1966Neutron}.

Notably, spin transport phenomena in bilayer films of atomic layer materials, including graphene, and magnets have been theoretically predicted in spintronics, suggesting potential contributions to the study of transport phenomena in atomic layer materials. In this context, analyzing SSE in similar systems through microscopic theory could provide an opportunity to apply the extensive knowledge gained from spin caloritronics to atomic layer material research.
Focusing on the vicinity of the interface, SSE and SP share similarities in their spin current generation mechanisms, which suggests that their theoretical analysis methods are closely related. However, there is a key difference in the driving forces: SSE is driven thermally, while SP operates coherently at ferromagnetic resonance frequencies, leading to distinct spin transport characteristics.
The aim of this research is to forge a novel path in spin caloritronics using atomic layer materials. The investigation starts with an analysis of SSE in a graphene/magnet bilayer system. 

One of the notable features of graphene is the significantly larger Landau-level separations compared to the two-dimensional electron gas in conventional semiconductor heterojunctions \cite{Novoselov2005,zhang2005experimental,novoselov2007room,neto2009electronic}. This allows for the observation of Landau quantization at relatively high temperatures and weak magnetic fields. The reported SSE experiments have been conducted in magnetic fields exceeding $10$ T, which sufficiently meet the practical conditions for observing the effects of Landau quantization in graphene. Therefore, it is important to elucidate the effect of Landau quantization on the SSE in a graphene/magnet bilayer system.
This paper aims to lay the groundwork for pioneering research in spin caloritronics with atomic layer materials employing microscopic theory.

This paper is organized as follows. The model for the graphene/ferromagnetic insulator (FI) interface is introduced in Sec.~\ref{system Hamiltonian}.
The properties of the dynamic spin susceptibilities of graphene with spin-splitting and the FI are summarized in Sec.~\ref{dynamic}.
The microscopic expressions for the tunneling spin currents at the interface generated by the SSE are given by using the Schwinger-Keldysh approach in Sec.~\ref{SC}.
The numerical analysis of the spin currents is shown in Sec.~\ref{neu sec}. We reveal that the spin currents exhibit quantum oscillations originating from the Landau quantization. 
We compare the spin current caused by the SSE with that cause by SP in the quantum Hall regime of graphene.
We propose an experimental setup for the detection of the SSE at the graphene/FI interface in Sec.~\ref{detection}.
Finally, our results are summarized in Sec.~\ref{sum}. 

\section{System Hamiltonian }\label{system Hamiltonian}

\begin{figure}[htbp]
    \centering
    \includegraphics[width=0.5\textwidth]{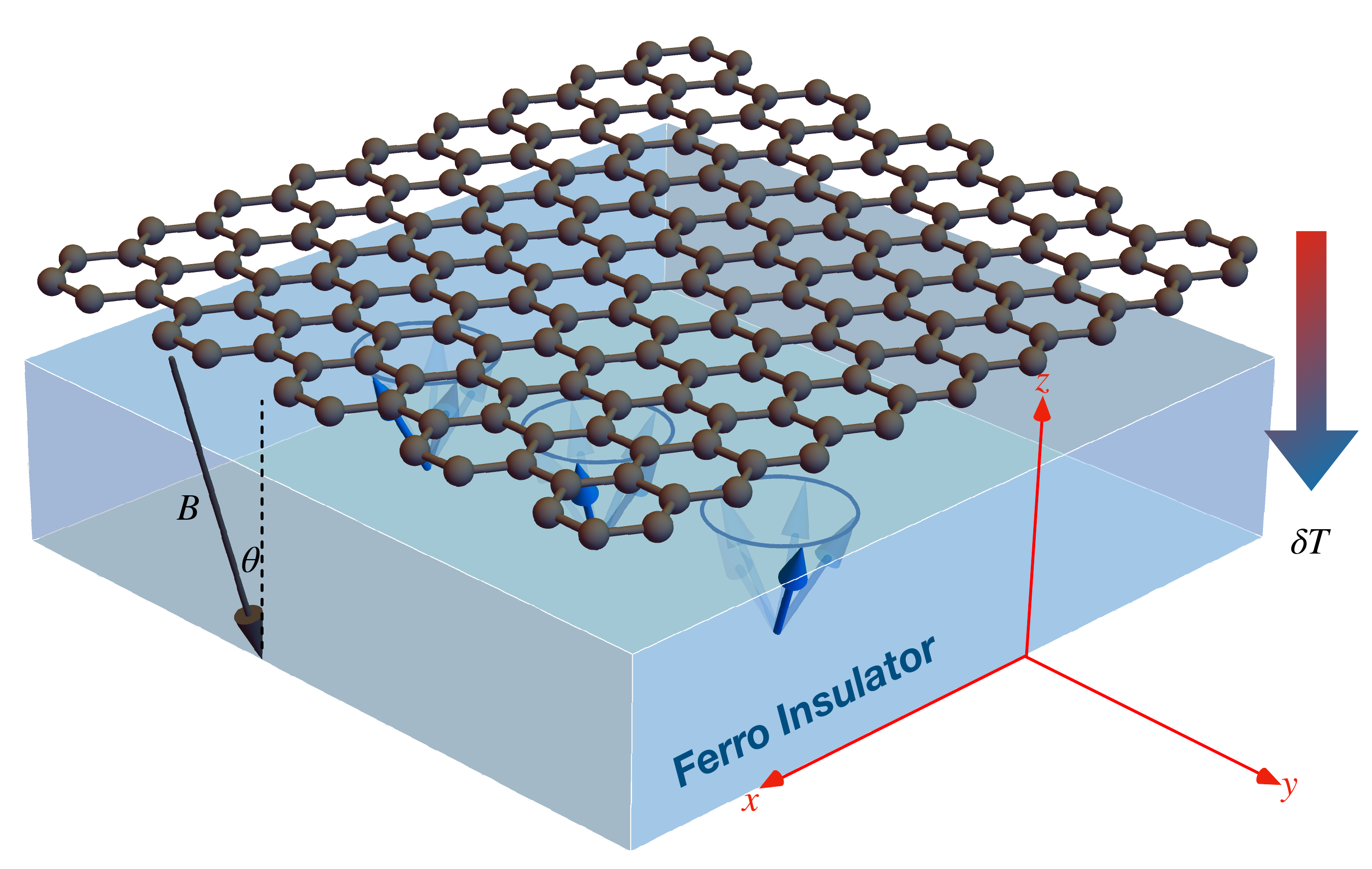}
    \caption{A schematic picture of the setup considered in this article. A ferromagnetic insulator is placed on a graphene monolayer and subjected to an external magnetic field $B$ with tilted angle $\theta$. A temperature bias $\delta T$ is applied across this bilayer, activating spin tunneling transport at the interface caused by the spin Seebeck effect.}
    \label{system}
\end{figure}

Figure~\ref{system} shows a schematic picture of the bilayer system composed of a graphene monolayer and a FI.
The external magnetic field is tilted by $\theta$ from the perpendicular direction, which is important to detect the SSE (See Sec.~\ref{detection}).
The $z$-component of the Zeeman field discretizes the electronic states in graphene into Landau levels through a non-perturbative magnetic effect in the regime of sufficiently strong magnetic fields. The spin polarization of the tunneling spin current generated by a thermal gradient aligns with the Zeeman field. Thus, when the magnetic field is along the $z$-axis, the spin polarization also points along $z$-axis, making direct detection via the inverse spin Hall effect (ISHE) impractical, as ISHE requires an in-plane spin polarization for transverse charge signals. Tilting the Zeeman field away from the $z$-axis, as shown in Fig.~\ref{system}, introduces an in-plane spin component, enabling ISHE detection.
In the following calculation, we set $\theta=0$. The results for $\theta\neq0$ are obtained by replacing $B$ with $B\cos\theta$.

The system Hamiltonian is given by
\begin{align}
    H = H_{\mathrm{G}} + \HFI + H_{\mathrm{ex}}.
\end{align}
The first term $H_{\mathrm{G}}$ describes the electronic states in graphene and is given by
\begin{align}
    H_{\mathrm{G}}=\sum_p\varepsilon_pc_p^\dagger c_p,
\end{align}
where $c_p^\dagger$ ($c_p$) denotes the creation (annihilation) operator with a set of quantum number $p$ and eigenenergy $\varepsilon_p$.
The low energy electronic states in graphene are well described by the effective Hamiltonian around the $K$ and $K^\prime$ points \cite{neto2009electronic}.
The eigenenergy and eigenstates around the $K$ point are obtained by diagonalizing the effective Hamiltonian
\begin{align}
    \mathcal{H}_{\mathrm{eff}}=v\mqty(0 & \pi_x-i\pi_y \\ \pi_x+i\pi_y & 0),
    \label{eq_graphene_eff}
\end{align}
where $v\approx10^6$ $\mathrm{m/s}$ is the Fermi velocity and $\vb*{\pi}=-i\hbar\nabla+e\vb*{A}$ with the vector potential $\vb*{A}$.
In this paper, we assume each valley can be treated independently so that the valley degeneracy just doubles the tunneling spin current.

The second term $\HFI$ describes the bulk FI and is given by
\begin{align}
    \HFI=-\mathcal{J}\sum_{\langle i,j\rangle}\vb*{S}_i\cdot\vb*{S}_j-\hbar\gamma B\sum_j S^z_j,
\end{align}
where $\vb*{S}_j$ is the localized spin in the FI at site $j$, $\mathcal{J}$ is the coupling constant, $\langle i,j\rangle$ means summation for nearest neighbors, and $\gamma$ is the gyromagnetic ratio.
Using the Holstein-Primakoff transformation \cite{PhysRev.58.1098} and employing the spin-wave approximation ($S_j^z=S-b_j^\dagger b_j$, $S_j^+\approx\sqrt{2S}b_j$), $\HFI$ is rewritten as
\begin{align}\label{HFI}
    \HFI\approx\sum_{\vb*{k}}\hbar\omega_{\vb*{k}}b_{\vb*{k}}^\dagger b_{\vb*{k}},
\end{align}
where $\hbar\omega_{\vb*{k}}$ is the magnon dispersion, $b_{\vb*{k}}^{\dagger}$ ($b_{\vb*{k}}$) denotes the creation (annihilation) operator of magnons with the wave vector $\vb*{k}$. Here, we have omitted a constant term. For simplicity, we assume that $\hbar\omega_{\vb*{k}}$ is given by 
\begin{align}\label{DIS}
    \hbar\omega_{\vb*{k}}=2\mathcal{J}Sa^2k^2+\hbar\gamma B ,
\end{align}
where $a$ is the lattice constant of the FI.

The third term $\Hex$ describes the proximity exchange coupling at the interface and is given by
\begin{align}
    &\Hex
    =-\int d\vb*{r}\sum_nJ(\vb*{r},\vb*{r}_j)\vb*{s}(\vb*{r})\cdot\vb*{S}_j
    =H_Z+H_T, \\
    &H_Z=-\int d\vb*{r}\sum_jJ(\vb*{r},\vb*{r}_j)s^z(\vb*{r})S^z_j, \\
    &H_T=-\frac{1}{2}\int d\vb*{r}\sum_j
    J(\vb*{r},\vb*{r}_j)
    \qty(s^+(\vb*{r})S^-_j+s^-(\vb*{r})S^+_j).
\end{align}
$\vb*{s}(\vb*{r}):=\sum_{p,q}[\phi_p(\vb*{r})c_p]^\dagger\vb*{s}\phi_q(\vb*{r})c_q$ is the spin density operator in graphene, where $\phi_p(\vb*{r})$ is the eigenstate of Eq.~(\ref{eq_graphene_eff}) and $\vb*{s}$ are the Pauli matrices in the spin space.
Here, we assume that $H_Z$ is approximated as
\begin{align}
    &H_Z\approx -J_0Ss^z_{\mathrm{tot}}, \\
    \label{Zeemman}
    &s^z_{\mathrm{tot}}=\int d\vb*{r}s^z(\vb*{r}).
\end{align}
In our system, $H_Z$ leads to the spin splitting in graphene and $H_T$ describes the spin transfer at the interface.

We treat $H_{\mathrm{G}}+\HFI+H_Z$ as an unperturbed Hamiltonian and $H_T$ as a perturbation in the following calculation on the tunneling spin current.
Then, the spin-split Landau levels in graphene are given by
\begin{align}
    \varepsilon_{ns}=\mathrm{sgn}(n)\sqrt{2e\hbar v^2}\sqrt{|n|B}-J_0Ss, \label{eq_ssLL}
\end{align}
with the Landau level index $n=0,\pm1,\pm2\cdots$ and the spin index $s=\pm$.
The energy bands in the absence of the magnetic field are given by
\begin{align}
    \varepsilon_{n\vb*{k}s}=n\hbar vk-J_0Ss, \label{eq_band}
\end{align}
where $n=\pm 1$, the positive and negative sign denote conduction and valence bands, respectively.
We will later confirm that the tunneling spin current in the $B\to0$ limit is consistent with the tunneling spin current in the absence of the magnetic field.

\section{Dynamic spin susceptibility}\label{dynamic}

In this section, we summarize the properties of the dynamic spin susceptibilities of graphene with spin-splitting and the FI, which are later used in the expression of the tunneling spin current.
The dynamic spin susceptibility of graphene is defined as
\begin{align}
    \chi^R(\vb*{r},\omega)
    :=\int dte^{i(\omega+i0)t}\frac{i}{\hbar}
    \theta(t)\langle[s^+(\vb*{r},t),s^-(\vb*{0},0)]\rangle_0 ,
\end{align}
where the average $\langle \cdots \rangle _0$ is taken for the unperturbed Hamiltonian while $\langle \cdots \rangle$ is taken for the full Hamiltonian as introduced later. We define the local spin susceptibility as
\begin{align}
    \chi_{\mathrm{loc}}^R(\omega):=\chi^R(\vb*{0},\omega).
\end{align}
Using the eigenstates and the eigenenergy of Eq.~(\ref{eq_graphene_eff}), we obtain $\mathrm{Im}\chi_{\mathrm{loc}}^R(\omega)$
\begin{align}\label{IMO}
    &\mathrm{Im}\chi^R_{\mathrm{loc}}(\omega) \notag \\
    &=
    \frac{\pi}{2}\int d\varepsilon
    \qty[ f_{\mathrm{FD}}(\varepsilon)-f_{\mathrm{FD}}(\varepsilon+\hbar\omega)]
    D^\alpha_+(\varepsilon)D^\alpha_-(\varepsilon+\hbar\omega),
\end{align}
where $f_{\mathrm{FD}}(\varepsilon)=1/(e^{(\varepsilon-\mu)/k_BT}+1)$ is the Fermi distribution function with the chemical potential of graphene $\mu$. 
Additionally, $D_s^{\mathrm{LL}}(\varepsilon)$ and $D_s^{\mathrm{PW}}(\varepsilon)$ are the density of states with spin index $s$ in the presence and absence of the magnetic field, respectively.
The former $D_s^{\mathrm{LL}}$ is given by
\begin{align}
    D^{\mathrm{LL}}_s(\varepsilon)
    =
    \frac{1}{2\pi\ell_{\mathrm{B}}^2}\sum_n
    \frac{1}{\pi}
    \frac{\Gamma}{(\varepsilon-\varepsilon_{ns})^2+\Gamma^2},
\end{align}
where $\ell_{\mathrm{B}}=\sqrt{\hbar/(e B)}$ is the magnetic length and we have introduced a constant $\Gamma$ describing the Landau level broadening. 
The latter $D_s^{\mathrm{PW}}(\varepsilon)$ is given by
\begin{align}
    D^{\mathrm{PW}}_s(\varepsilon)
    =
    \frac{1}{2\pi(\hbar v)^2}\abs{\varepsilon-J_0Ss}.
\end{align}

The dynamic spin susceptibility of the FI is defined as
\begin{align}
    G^R(\vb*{k},\omega)
    :=\int dte^{i(\omega+i0)t}
    \frac{1}{i\hbar}\theta(t)
    \langle[S_{\vb*{k}}^+(t),S_{-\vb*{k}}^-(0)]\rangle_0,
\end{align}
which is calculated as
\begin{align}
    G^{R}(\vb*{k},\omega)
    =
    \frac{2S}{\hbar}
    \frac{1}{\omega-\omega_{\vb*{k}}+i\alpha\omega},
\end{align}
where we have introduced the phenomenological damping parameter $\alpha$.
The local spin susceptibility of the FI is defined as
\begin{align}
    G^R_{\mathrm{loc}}(\omega)
    :=\frac{1}{N}\sum_{\vb*{k}}G^R(\vb*{k},\omega).
\end{align}
The density of states of magnons per sites is defined as
\begin{align}
    D_m(\varepsilon):=\frac{1}{N}\sum_{\vb*{k}}\delta(\varepsilon-\hbar\omega_{\vb*{k}}).
\end{align}
Assuming $\alpha\to0$, the density of states is given by
\begin{align}
    D_m(\varepsilon)=-\frac{1}{2\pi S}\mathrm{Im}G^R_{\mathrm{loc}}(\varepsilon/\hbar),
\end{align}
which is explicitly calculated as
\begin{align}
    D_m(\varepsilon)=\frac{(2J_0S)^{-3/2}}{4\pi^2}\sqrt{\varepsilon-\hbar\gamma B}.
\end{align}

\section{Tunneling Spin Current}\label{SC}

The tunneling spin current operator is defined as
\begin{align}\label{Is}
    I_s^z:=\frac{i}{2}[s^z_{\mathrm{tot}},H].
\end{align}
Substituting the total Hamiltonian and using the commutation relation of the spin density operators, we obtain
\begin{align}
    I_s^z
    =
    -\frac{i}{2}\int d\vb*{r}
    \sum_j
    [J(\vb*{r},\vb*{r}_j)s^+(\vb*{r})S_j^- - \mathrm{H.c.}].
\end{align}
We calculate the statistical average of $I_s^z$ within the second order perturbation calculation. We assume a nonequilibrium steady state with the temperature difference between graphene and the FI. Consequently, the tunneling spin current is given by
\begin{align}\label{general}
    \langle I_s^z\rangle
    =
    2\hbar J_2^2l^2A\int \frac{d\omega}{2\pi}
    &\mathrm{Im}\chi^R_{\mathrm{loc}}(\omega)
    \qty[-\mathrm{Im}G^R_{\mathrm{loc}}(\omega)] \notag \\
    &\times\qty(
        \frac{\partial f_{\mathrm{BE}}(\omega,T)}{\partial T} 
    )\delta T,
\end{align}
where we added a factor 2 considering the contribution of both valleys, $f_{\mathrm{BE}}(\omega,T)=1/(e^{\hbar\omega/k_{\mathrm{B}}T}-1)$ is the Bose-Einstein distribution function, $T=T_{\mathrm{G}}$ is the temperature of graphene, and $\delta T=T_{\mathrm{FI}}-T_{\mathrm{G}}$ is the temperature difference.
Note that we have used the following replacement
\begin{align}\label{approx}
    \sum_{j,j^\prime}
    J(\vb*{r},\vb*{r}_j)
    J(\vb*{r}^\prime,\vb*{r}_{j^\prime})
    e^{-i(\vb*{r}_j-\vb*{r}_{j^\prime})\cdot\vb*{k}}
    \to
    J_2^2l^2\delta(\vb*{r}-\vb*{r}^\prime),
\end{align}
where $J_2^2l^2$ is the variance with a characteristic energy $J_2$ and a characteristic correlation length $l$.
The above replacement corresponds to taking interface configuration average \cite{Ominato2022-se, Ominato2022-xc}, which reproduces the expression in previous work on the SSE at the interface of the SC/FI bilayer system \cite{Kato2019-zf}.
We introduce a characteristic spin current unit
\begin{align}\label{factor}
    I_0^z
    =
    \frac{2J_2^2l^2Ak_{\mathrm{B}}\delta T}{\mathcal{J}\sqrt{\mathcal{J}S}}
    \frac{10^{-9}}{\sqrt{\mathrm{meV}}\mathrm{nm}^4},
\end{align}
which we use for normalization to plot the numerical results as shown below.

\section{Numerical results}\label{neu sec}
We numerically examine the tunneling spin current generated by the SSE.
In the following, we analyze dimensionless spin currents $I_n^\alpha$ defined as
\begin{align}
    I_n^\alpha:=\frac{\langle I_{s}^{z}\rangle_{\alpha}}{I^z_0}\hspace{2mm}(\alpha = \mathrm{LL},~\mathrm{PW}),
\end{align}
where $\langle I_{s}^{z}\rangle_{\mathrm{LL}}$ and $\langle I_{s}^{z}\rangle_{\mathrm{PW}}$ represent the spin current in the presence and absence of the magnetic field, respectively. During the numerical calculation, for the integral of frequency $\omega$ in Eq.~(\ref{general}), we chose $E_M=10$ meV as the high-energy cut-off of the magnon dispersion relation, which is of the order of the exchange interaction in the FI.

\begin{figure}[htbp]
    \centering
    \includegraphics[width=1\hsize]{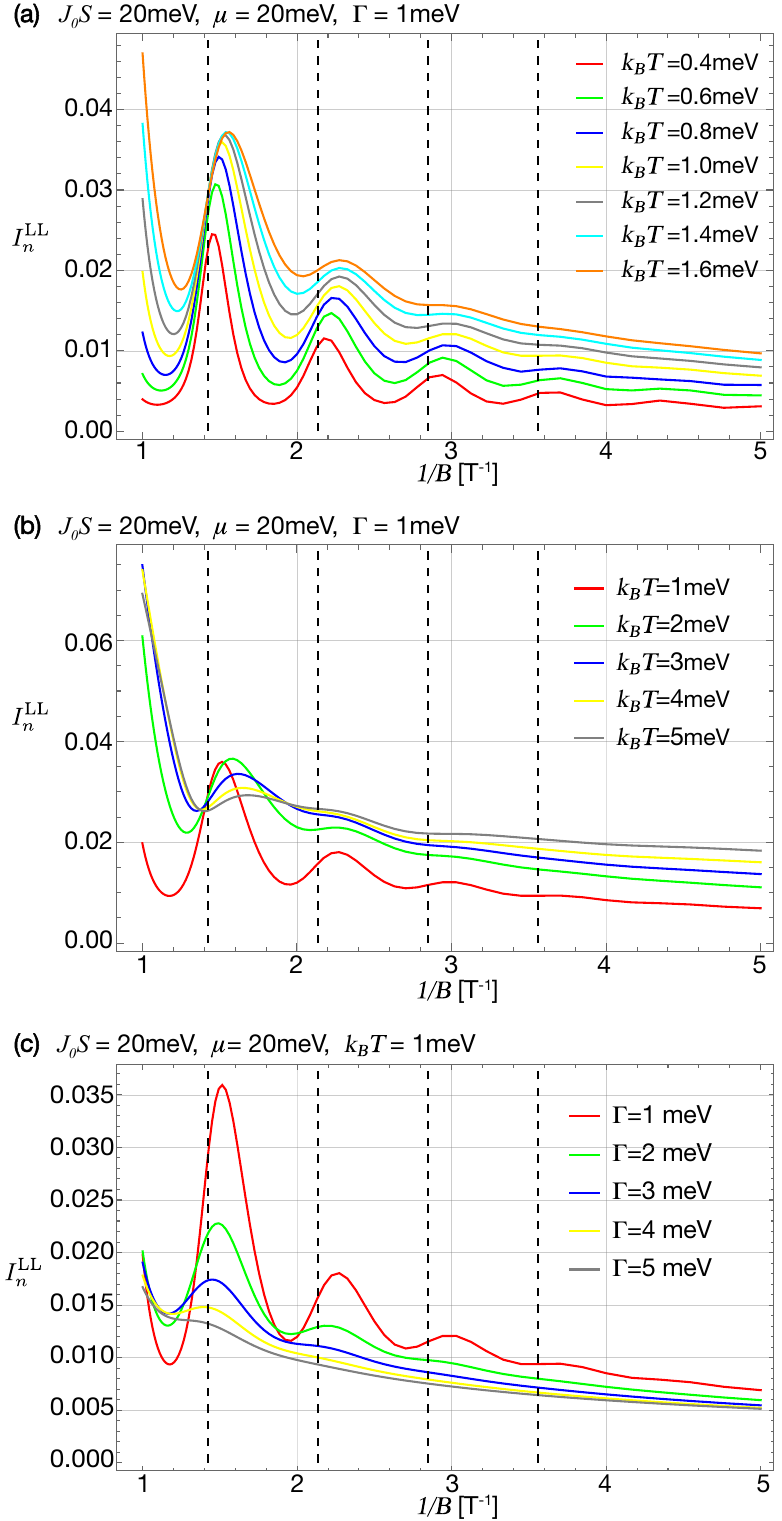}
    \caption{(a) (b) The dimensionless spin current $I_n^{\mathrm{LL}}$ as a function of $1/B$ for different temperatures. (c) $I_n^{\mathrm{LL}}$ as a function of $1/B$ for different Landau level broadening $\Gamma$. The vertical dashed lines indicate the peak positions evaluated by the crossing points of the up and down spin Landau levels (see the main text).}
    \label{draft1}
\end{figure}

\begin{figure}[htbp]
    \centering
    \includegraphics[width=1\hsize]{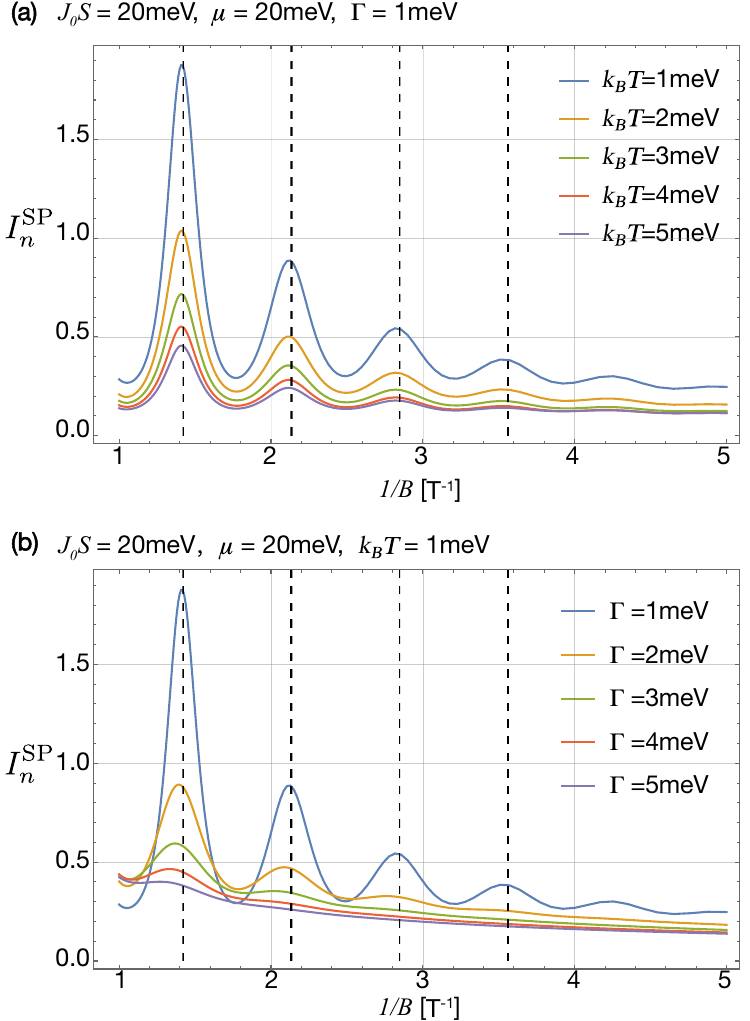}
    \caption{The tunneling spin current generated by the SP. The dimensionless spin current as a function of $1/B$ (a) for several temperatures and (b) for several level broadening $\Gamma$'s. The vertical dashed lines indicate the peak positions evaluated by the crossing points of the up and down spin Landau levels (see the main text).}
    \label{draft2}
\end{figure}

Figure~\ref{draft1}  shows the tunneling spin current $I_n^{\mathrm{LL}}$ as a function of the inverse of the magnetic field $1/B$. Figures~\ref{draft1} (a) and (b) show the plots for various temperatures with fixed values of spin splitting $J_0S$, chemical potential $\mu$, and level broadening $\Gamma$. The spin current oscillates as a function of $1/B$, and the oscillation is suppressed with the increase of the temperature. This is a quantum oscillation originating from Landau quantization, indicating that the electronic system is in the quantum Hall regime. Figure~\ref{draft1} (c) shows the spin current for several level broadening with fixed spin splitting $J_0S$, chemical potential $\mu$, and temperature $T$. Similar to the effect of increasing temperature, the quantum oscillation is suppressed as the level broadening increases. A similar quantum oscillation phenomenon has been theoretically reported in SP\cite{ominato2020quantum}. We have reproduced the quantum oscillations that arise owing to SP, as shown in Fig.~\ref{draft2}, and confirmed the similar suppression of the quantum oscillation with the increase in temperature and level broadening.

Figures~\ref{draft1} and \ref{draft2} also represent
two differences between SSE and SP as the temperature increases. First, the peak positions of the spin current generated by SSE shift to the right, while the peak positions of the spin current generated by SP are independent of temperature (see Appendix~\ref{PeakPositionsSP} for details).
Second, the spin current generated by SSE shows an increasing tendency contrary to the spin current generated by SP showing a decreasing tendency. These are because the SSE is driven by the temperature gradient, which allows higher frequency magnons to contribute to the spin current.

\begin{figure*}[htbp]
    \centering
    \includegraphics[width=1\textwidth]{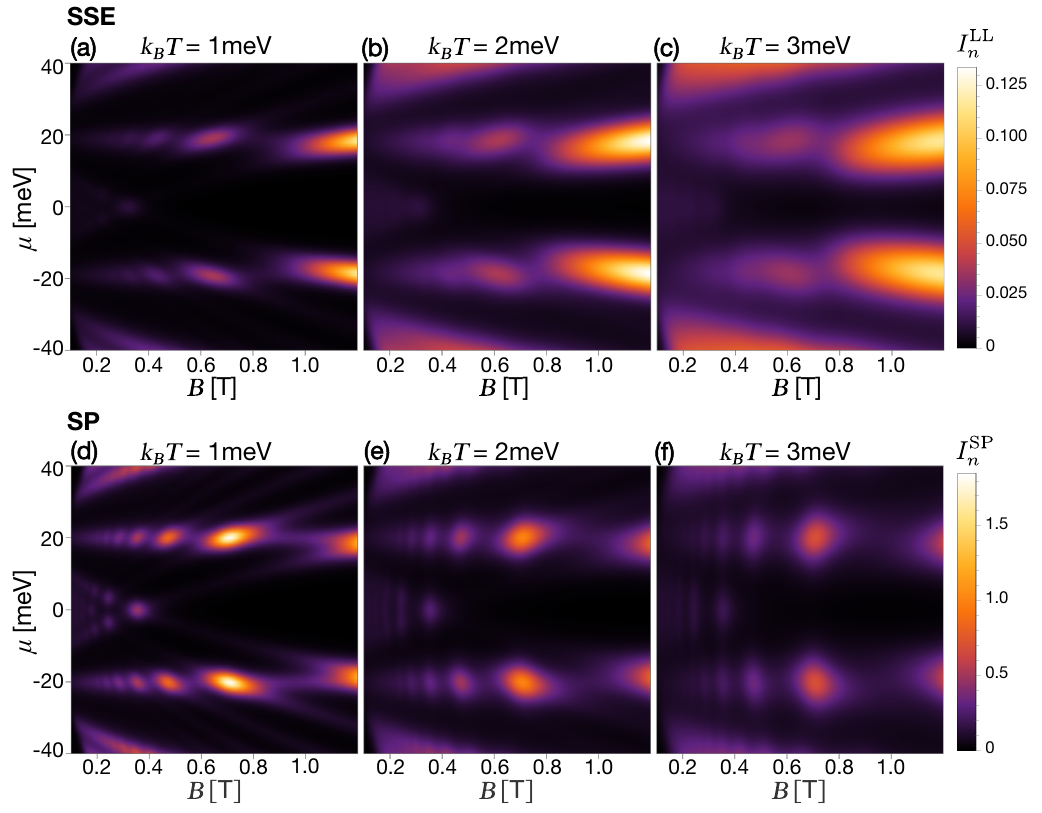}
    \caption{The density plots of the dimensionless spin current generated by the SSE (top panels (a)-(c)) and the SP (bottom panels (d)-(f)). We set $\mu=J_0S=20~\mathrm{meV}$ and $\Gamma=1~\mathrm{meV}$.}
    \label{draft3}
\end{figure*}

\begin{figure*}[htbp]
    \centering
    \includegraphics[width=1\textwidth]{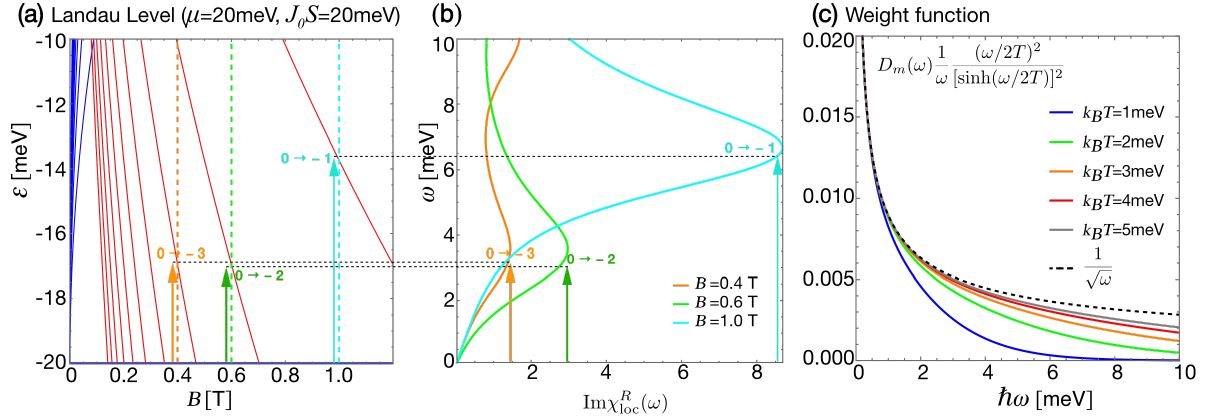}
    \caption{We set the spin-splitting energy $J_0S=20~\mathrm{meV}$ and chemical potential $\mu=20~{\mathrm{meV}}$. (a) The Landau level structure and possible excitation processes with spin flipping for the different magnetic field $B$, where $0 \rightarrow -1$ means the excitation process from  $\varepsilon_{0,+}$ to $\varepsilon_{-1,-}$. (b) The dynamic spin susceptibility $\mathrm{Im}\chi^R_{\mathrm{loc}}(\omega) $ as a function of $\omega$ and (c) the weight function defined as $W(\omega,T)= D_m(\omega)\frac{1}{\omega}\frac{(\omega/2T)^2}{[\sinh(\omega/2T)]^2}$ as a function of $\omega$ for the several temperature $T$. The dasshed currve represents $1/\sqrt{\omega}$.}
    \label{draft4}
\end{figure*}

In Fig.~\ref{draft3}, we compare the SSE and SP signals, including their chemical potential dependence, using density plots.  
Both of their bright regions originate from the crossing points of spin-up and spin-down Landau levels. 
Since the SP signal is only contributed by zero-frequency magnons, the bright spot does not spread; on the other hand, the SSE signal is broadened in the $\mu$ direction owing to the contribution of thermally excited higher-frequency magnons, and the peak itself grows as the temperature increases. 
The representative spin-flipping processes that lead to such broadening are shown in Figs.~\ref{draft4} (a) and (b). The thermal broadening of the SSE and the increase in the signal at higher temperatures are attributed to the behavior of the weight function defined by
\begin{align}
    W(\omega,T)=D_m(\omega) \frac{\partial f_{BE}}{\partial T},     
\end{align}
as illustrated in Fig.~\ref{draft4} (c).

We emphasize the significance of the quantum oscillation of the tunneling spin current in the SSE indicated above.
Attempts have been made to detect the unique properties of quantum states through spin tunneling transport across magnetic interfaces.
For example, in $s$-wave superconductor/ferromagnet bilayers, the coherence peak of the superconductor is visible through SP, while it has been reported that it is not seen in the SSE \cite{Kato2019-zf}. This is attributed to the fact that, although the SSE and SP are similar in terms of tunneling spin transport at magnetic interfaces, the SSE examines a thermal, or incoherent, spin response, whereas SP reveals a coherent spin response because of ferromagnetic resonance under microwave irradiation.
From this difference, careful consideration is required to determine whether both methods can detect information about quantum states through interface spin currents. Under such premises, the results obtained in Fig.~\ref{draft1} present a noteworthy outcome: Unlike the coherence peak of the superconductor, quantum oscillations characteristic of the quantum Hall regime are detectable in both the SSE and SP \cite{ominato2020quantum}.

\begin{figure}[htbp]
    \centering
    \includegraphics[width=0.5\textwidth]{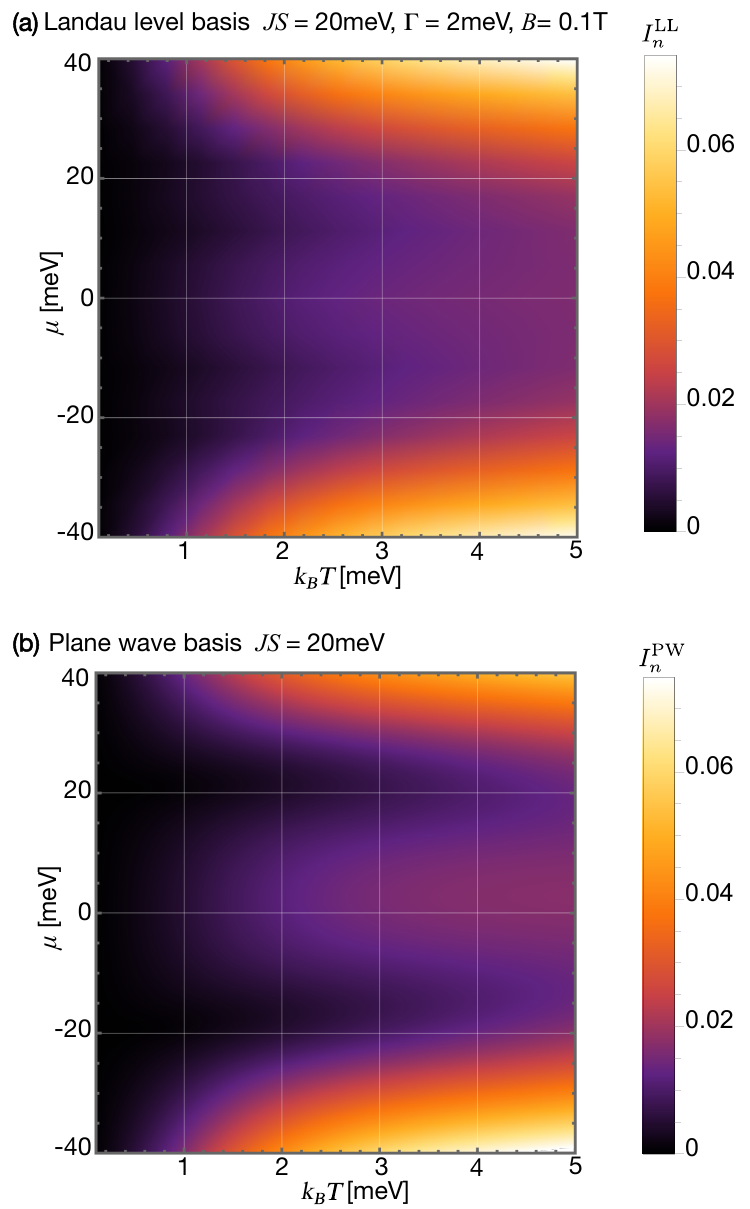}
    \caption{
    The density plot of the dimensionless spin current (a) $I_n^{\mathrm{LL}}$ and (b) $I_n^{\mathrm{PW}}$ as a function of $\mu$ and $k_BT$. We set $B=0.1~\mathrm{T}$ and $\Gamma=2~\mathrm{meV}$ for the calculation of $I_n^{\mathrm{LL}}$. The spin splitting $J_0S$ is set to $20~\mathrm{meV}$ in both cases.}
    \label{P3}
\end{figure}

Finally, we compare the spin current owing to the SSE calculated in the weak magnetic field limit using the Landau level basis with the spin current calculated using the plane wave basis, and confirm that they are consistent. Here, the weak magnetic field limit refers to a condition where the Landau level spacing is much smaller than both the level broadening $\Gamma$ and the thermal energy $k_BT$. Figures \ref{P3}(a) and \ref{P3}(b) show the density plots of the spin current as a function of $\mu$ and $k_BT$, calculated using the Landau level basis and the plane wave basis, respectively. Their qualitative behaviors are similar to each other. In contrast to the strong magnetic field region, where the effects of Landau quantization are significant, there is almost no spin current generation at the energy corresponding to the Dirac point in the weak magnetic field limit. This is because, the density of states at the Dirac point becomes zero in the weak magnetic field limit, while the density of states at the zeroth Landau level significantly contributes to the spin current in the strong magnetic field region.

\section{Detection}
\label{detection}

\begin{figure}[htbp]
    \includegraphics[width=1\hsize]{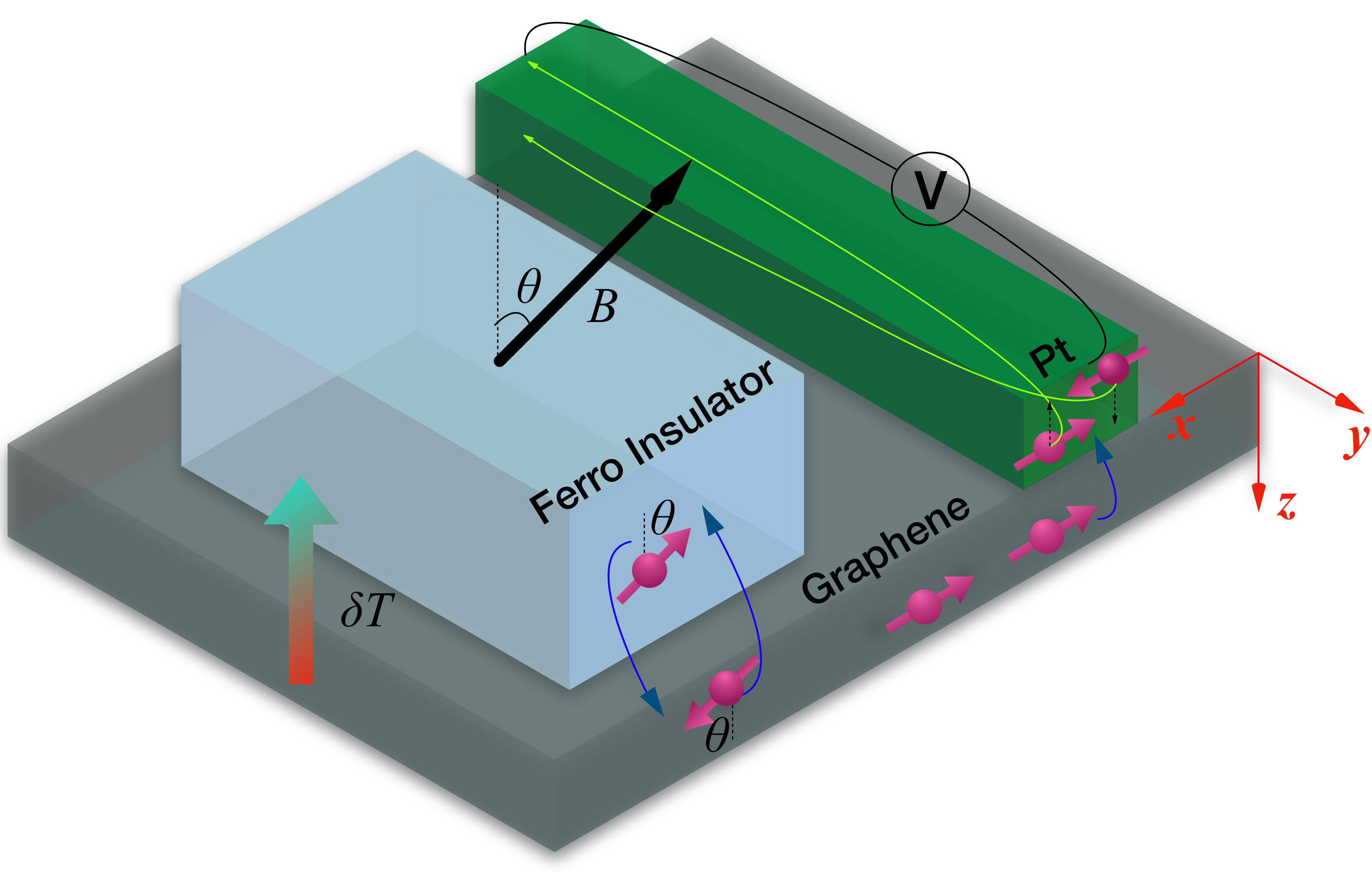}
    \caption{An experimental setup for electrically detecting the spin current generated by the SSE. When a temperature gradient is generated between the FI and graphene, a spin current is generated at the interface, which gives the diffusive spin current in the graphene. The diffusive spin current is injected into platinum through the interface between platinum and graphene. When the injected spins have a finite $x$ component, a voltage is generated in the $y$ direction of the platinum because of the inverse spin Hall effect.} 
    \label{exp}
\end{figure}

We discuss the detection methods of the SSE at the graphene/FI interface. Here, we propose an electrical measurement method using the inverse spin Hall effect (ISHE) as shown in Fig.~\ref{exp}. By tilting the applied external magnetic field from the surface normal direction by an angle $\theta$, the surface normal component of the magnetic field generates a spin current because of the SSE. The component of the spin current in the horizontal direction will be transported into the Pt layer and the spin current is converted to the electric potential through the ISHE. The SP from a magnet to a graphene monolayer were observed in a similar experimental setup \cite{PhysRevLett.116.166102, PhysRevB.87.140401, PhysRevApplied.10.044053}.

\section{Discussion}
Lastly, we note the relevance of spin-orbit coupling (SOC) in the context of spin transport. In this study, we investigated tunneling spin transport at the interface between a ferromagnetic insulator and graphene, driven by a temperature gradient via the spin Seebeck effect. The tunneling processes and resulting spin currents are primarily governed by spin-conserving mechanisms. Because of the weak intrinsic SOC of graphene and the long spin relaxation times associated with it \cite{tombros2007electronic, han2010tunneling, yang2011observation, han2011spin, dlubak2012highly, han2014graphene}, the effects of SOC on interfacial tunneling are negligible. This justifies the omission of SOC in our theoretical model without compromising the validity of our main conclusions.

For spin current detection, we employed Pt becuase of its strong SOC, which facilitates efficient spin-to-charge conversion via the ISHE. While graphene with sufficiently strong SOC could, in principle, serve as an effective ISHE detector, the weak intrinsic SOC of graphene makes it unsuitable for this role. The use of Pt reflects a practical design choice to ensure reliable detection.
Although SOC is significant in bulk spin transport phenomena, its influence on the interfacial tunneling processes analyzed here is minimal. Nevertheless, proximity-induced SOC in graphene through adjacent materials with strong SOC represents a promising avenue for future exploration\cite{avsar2014spin, wang2015strong, gmitra2015graphene, wang2016origin, yang2016tunable, dankert2017electrical, ghiasi2017large, gmitra2017proximity, yang2017strong, volkl2017magnetotransport, zihlmann2018large, wakamura2018strong, benitez2018strongly}. Such effects could potentially enable graphene to function as an efficient ISHE detector, broadening its applicability in spintronic devices and reducing the need for external materials such as Pt. This perspective highlights the potential for expanding the role of graphene in advanced spintronic technologies.

\section{Summary}\label{sum}
We have developed a microscopic theory to describe the SSE at the magnetic interface of graphene/ferromagnetic bilayers, focusing on a detailed comparison of the spin current generated by both SSE and SP.
We found that both exhibit quantum oscillations, with SSE showing a peak shift relative to SP owing to the contribution of higher frequency magnons.
In particular, we revealed that thermally excited higher frequency magnons significantly contribute to the SSE.
Our theory provide useful insights into the manipulation and control of spin currents in spincaloritronic devices, marking a significant step forward in our understanding of spin-thermal interconversion at magnetic interfaces consisting of atomic later materials.

\section{Acknowledgement}
This work was supported by the National Natural Science Foundation of China (NSFC) under Grant No. 12374126, 
by the Priority Program of the Chinese Academy of Sciences under Grant No. XDB28000000, and by JSPS KAKENHI under Grants (No.21H01800, No.21H04565, No.23H01839, and No.24H00322) from MEXT, Japan. 

\appendix
\section{The eigenstates of the graphene}\label{A1}
In this appendix, we show the eigenstates of graphene used in the calculation of this article with and without the external magnetic field $B$.

When an external magnetic field $B$, the electrons settle in Landau energy levels in graphene,
\begin{align}\label{eLL}
    \varepsilon_{ns}=\mathrm{sgn}(n)\sqrt{2e\hbar v^2}\sqrt{|n|B}-J_0S s, 
\end{align} 
where $n=0,\pm1,\pm2\cdots$ denotes the Landau level index, $s=\pm$ is the spin index and $v\approx10^6$ $\mathrm{m/s}$ is the Fermi velocity. The last term is the spin-splitting energy originating from $H_Z$ as one part of the interfacial interaction $H_\mathrm{int}$.
With the Landau gauge, which gives $\boldsymbol{A}=(0,Bx,0)$ and $B$ is the pure magnetic field in the z direction, the corresponding eigenstates are:
\begin{align}\label{LE1}
    \phi_{0X}(\vb*{r})&=\frac{e^{ik_yy}}{\sqrt{L_y}}\begin{pmatrix} 0 \\ u_{0}(x-X) \end{pmatrix},\\
    \label{LE2} \phi_{nX}(\vb*{r})&=\frac{e^{ik_yy}}{\sqrt{2L_y}}\begin{pmatrix} \mathrm{sgn}(n)u_{|n|-1}(x-X) \\ u_{|n|}(x-X) \end{pmatrix},
\end{align}
with 
\begin{align}
    u_n(x)=\frac{1}{\sqrt{2^nn!\sqrt{\pi}\ell_{\mathrm{B}}}}H_n(x/\ell_{\mathrm{B}})\exp(-x^2/2\ell_{\mathrm{B}}),
\end{align}
where $H_n(x/\ell_{\mathrm{B}})$ is the Hermite polynomials and $L_y$ is the length of the graphene sheet along the $y$-direction.

While the energy bands in the absence of the magnetic field are given by
\begin{align}\label{eP}
    \varepsilon_{n\vb*{k}s}=n \hbar v k-J_0Ss,
\end{align}
where $n=\pm$, the positive and negative signs denote conduction and valence bands, respectively. The corresponding eigenstates are :
\begin{align}\label{LP}
    \phi_{n\vb*k}(\vb*{r})=\frac{e^{i\vb*k\cdot \vb*{r}}}{\sqrt{2L_xL_y}}\begin{pmatrix} 1 \\ n e^{i\varphi_{\vb*k}}\end{pmatrix},
\end{align} 
where $\varphi_{\vb*k}=\operatorname{arctan}(k_y/k_x)$.

\section{The spin-wave approximation in FI}
Here we describe how the Holstein-Primakoff transportation\cite{PhysRev.58.1098} and the spin-wave approximation are used to define the magnon operators $b^{(\dagger)}_{\vb*{k}}$ in Eq.~(\ref{HFI}). Firstly, define $S^{(\pm)}_{\vb*{k}}$ by the Fourier transformation:
\begin{align}
    \label{sk_1}
    & S^+_j=\frac{1}{\sqrt{N}} \sum_{\vb*{k}} e^{i \vb*{k} \cdot \vb*{r}_j} S^+_{\vb*{k}},\\ 
    \label{sk_2}
    & S^-_j=\frac{1}{\sqrt{N}} \sum_{\vb*{k}} e^{-i \vb*{k} \cdot \vb*{r}_j} S^-_{-\vb*{k}}.
\end{align} 
where $N$ is the number of sites and $\vb*{k}$ is the wave vector of the spin-wave. Then we have:
\begin{align}
    S^+_{\vb*{k}} & \approx\sqrt{2S}b_{\vb*{k}}, \\
    S^-_{-\vb*{k}}& \approx \sqrt{2S}b^{\dagger}_{\vb*{k}}.
\end{align} 

\section{The interaction picture}
In this appenidx, we introduce the interaction picture and give the details of the Keldysh contour used in this article. In this article, the $H_T$ part of $H_{\mathrm{ex}}$ becomes the perturbation term Hamiltonian while the $H_{\mathrm{G}}+\HFI+H_Z$ is treated as the unperturbed Hamiltonian under the interaction picture as shown below. Since the transverse part of the Hamiltonian $H_T$ represents the spin transfer at the interface. This term is related to the distribution of $s(\vb*{r})$ and does not commute with the Hamiltonian of other parts of the system. All the operators appear with a time variable later means the operators under the interaction picture, for arbitrary operator ${O}$:
\begin{align}
    O(\vb*{r},t)=e^{iHt/\hbar}O(\vb*{r})e^{-iHt/\hbar},
\end{align}
where $H=H_{\mathrm{G}}+\HFI+H_Z$ as the unperturbed Hamiltonian. The operator $O(\vb*{r})$ is defined as: 
\begin{align}
    O(\vb*{r})
    = \sum_{p,q}[\phi_p(\vb*{r})c_p]^\dagger\vb*{O}\phi_q(\vb*{r})c_q,
\end{align} 
where $\vb*{O}$ is the corresponding matrix under Landau basis or plain wave basis, 
$\phi_p(\vb*{r})$ is the eigenstate of Eq.~(\ref{eq_graphene_eff}) which is discussed in detail in Appexdix.~\ref{A1} and $c^{(\dagger)}_q$ represents the annihilation (creation) operators of electrons in energy level $p$.
For spin density operators here, 
\begin{align}
    \vb*{s}(\vb*{r}):=\sum_{p,q}[\phi_p(\vb*{r})c_p]^\dagger\vb*{s}\phi_q(\vb*{r})c_q,
\end{align} 
where $\vb*{s}$ are the Pauli matrices in the spin space. 

The perturbation Hamilton $H_T$ inflects the evaluation of the quantum states:
\begin{align}
    |\Psi (t) \rangle &= U(t,-\infty)|\Psi (0) \rangle ,\\
    \langle \Psi (t)|&= \langle \Psi (0)|U^{\dagger}(t,-\infty),
\end{align}
where  
$U\left(t, t^{\prime}\right)=\mathcal{T}\exp \left(-\frac{i}{\hbar} \int_{t^{\prime}}^t H_{T}\left(t_1\right) \mathrm{d} t_1\right)$
and $\mathcal{T}$ is the time ordering operator. 
When considering the quantum expectation of arbitrary operator $O$, the Keldaysh contour is introduced as Fig.~\ref{f1}:
\begin{figure}[htbp]
    \centering
    \includegraphics[width=0.4\textwidth]{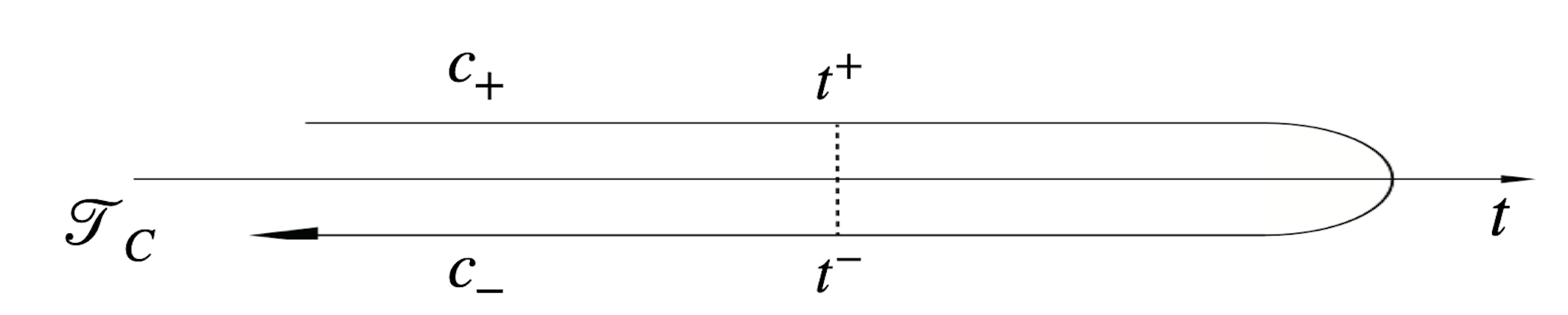}
    \caption{The Keldysh contour }
    \label{f1}
\end{figure}
\begin{align}
    \langle \Psi (t)|O|\Psi (t)\rangle &=\left\langle \Psi (0) \left| U^{\dagger}(t,-\infty)O(t)U(t,-\infty)\right| \Psi (0) \right\rangle \non \\
    &  =\langle \Psi (0)| U_{C} {O}(t)| \Psi (0) \rangle ,
   \end{align}
where $U_C=T_C \exp \left(-\frac{i}{\hbar} \int_C H_{T}\left(t_1\right) \mathrm{d} t_1\right)$ with the time ordering operator $T_{C}$ were redefined on Keldysh contour as Fig.~\ref{f1} shows. In this article, we use $\langle \cdots \rangle_0 $ to represent the quantum expectation under the initial density matrix (or the unperturbed Hamiltonian) and use $\langle \cdots \rangle $ to represent the average taken for the full Hamiltonian.

\section{Calculation of dynamic spin susceptibility}
In this appendix, we introduce the dynamic spin susceptibilities in graphene and FI to describe the spin current at the interface.
First, the most general propagators for the electron spin excitation in graphene and magnons in FI are defined as
\begin{align}\label{pro}
     \chi(\vb*{r},\vb*{r}';t,t')&:=\frac{i}{\hbar}\langle {T}_{C}{s}^{+}\left(\vb*{r} ,t\right){s}^{-} (\vb*{r} ',t')\rangle_0 ,\\
    G(\vb*{k};t,t')& :=\frac{1}{i\hbar}\langle {T}_{C}{S}_{\vb*{k}}^{+}(t) {S}_{-\vb*{k}}^{-}\left( t' \right) \rangle_0.
\end{align} 

\subsection{The Dynamic Spin Susceptibility in Graphene}
The dynamic spin susceptibility in graphene, also known as the retard component of the spin propagator, is defined as
\begin{align}\label{chi}
    \chi^R(\vb*{r},\vb*{r}';t,t'):=\frac{i}{\hbar}\theta(t)\langle[{s}^+(\vb*{r},t),{s}^-(\vb*{r}',t')]\rangle_0.
\end{align}
In a system with translational symmetry in both time and space, it reads
\begin{align}
        \chi^R(\vb*{r},\vb*{r}';t,t')&=\chi^R(\vb*{r}-\vb*{r}',\vb*{0};t-t',0) \non \\
        &=\frac{i}{\hbar}\theta(t)\langle[{s}^+(\vb*{r}-\vb*{r}',t-t'),{s}^-(\vb*{0},0)]\rangle_0,
    \end{align}
For this reason, only the retard component of the propagator starts from the zero point with $\vb*{r}=0$ and $t=0$ needed to be discussed, which is calculated as
    \begin{align}
     \chi^R(\vb*{r},\vb*{0};t,0)
    &=-\frac{1}{i\hbar}\theta(t)\langle[s^+(\vb*{r},t),s^-(\vb*{0},0)]\rangle_0 \non \\
    &=-\frac{1}{i\hbar}\theta(t)\sum_{j,k,l,m}
    \phi^\dagger_j(\vb*{r})
    \phi_k(\vb*{r})
    \phi^\dagger_l(\vb*{0})
    \phi_m(\vb*{0})
    s^+_{jk}s^-_{lm} \non \\
    &\quad \times e^{i(\varepsilon_j-\varepsilon_k)t/\hbar}
    \langle[c_j^\dagger c_k,c_l^\dagger c_m]\rangle_0 \non \\
    &=-\frac{1}{i\hbar}\theta(t)\sum_{j,k}
    \phi^\dagger_{j}(\vb*{r})\phi_{k}(\vb*{r})\phi^\dagger_{k}(\vb*{0})\phi_{j}(\vb*{0})s^+_{jk}s^-_{kj} \non \\
    & \quad \times e^{i(\varepsilon_j-\varepsilon_k)t/\hbar}(f_j-f_k),
    \end{align}
where the $\phi^{(\dagger)}(\vb*r)$, $c^{(\dagger)}$, $\varepsilon$ and $f$ is the eigen wave function, creation(annihilation) operator, energy and distribution function of electrons in a certain state. 
Here, we use $
    \langle[c_j^\dagger c_k,c_l^\dagger c_m]\rangle_0=\delta_{jm}\delta_{kl}(f_j-f_k)$.

Then, the local spin susceptibility is given by
\begin{align}
    \chi^R_{\mathrm{loc}}(t)&:=\chi^R(\vb*{0},\vb*{0};t,0) \non \\
    &=\frac{i}{\hbar}\theta(t)\sum_{j,k}|\phi_j^\dagger(\vb*{0})\phi_k(\vb*{0})|^2|s_{jk}^+|^2 \non \\
    &\quad \times  e^{i(\varepsilon_j-\varepsilon_k)t/\hbar}(f_j-f_k),
\end{align}
and the corresponding Fourier transformation $\chi^R_{\mathrm{loc}}(\omega)$ is given by
\begin{align}
    \chi^R_{\mathrm{loc}}(\omega)
    &=\int dte^{i(\omega+i0) t}\chi^R_{\mathrm{loc}}(t) \non \\
    &=-\sum_{j,k}|\phi_j^\dagger(\vb*{0})\phi_k(\vb*{0})|^2|s_{jk}^+|^2 \non \\
    &\quad \times \frac{f_j-f_k}{\varepsilon_j-\varepsilon_k+\hbar\omega+i0}.
\end{align}
The corresponding imaginary part is
\begin{align}\label{IMX}
    \mathrm{Im}\chi^{R}_{\mathrm{loc}}(\omega)
    &=\pi\sum_{j,k}|\phi_j^\dagger(\vb*{0})\phi_k(\vb*{0})|^2|s_{jk}^+|^2\delta(\varepsilon_j-\varepsilon_k+\hbar\omega)\non \\
    &\quad \times  \left[f_{\mathrm{FD}}(\varepsilon_j)-f_{\mathrm{FD}}(\varepsilon_j+\hbar\omega)\right].
\end{align}
 
Since we are discussing the electrons in graphene here, the $f_{i(j)}$ always be the Fermi-Dirac distribution function $f_{\mathrm{FD}}$ just as Eq.~(\ref{IMO}). We use $f_{\mathrm{FD}}$ to replace $f_{i(j)}$ here and in the following.

In the following, for the two different circumstances: the presence and absence of external magnetic field $B$, we calculate the $\mathrm{Im}\chi^{R}_{\mathrm{loc}}(\omega)$ separately. 

\subsubsection{Under Landau Level Basis}
When the eternal magnetic field $B$ exists, the eigenstates of the electrons with Landau gauge and the corresponding energy levels are already shown as Eqs.~(\ref{eLL}), (\ref{LE1}) and (\ref{LE2}). Recalling Eq.~(\ref{IMX}), the $\sum_{jk}$ represents $\sum_{X,X'}\sum_{n,n'}\sum_{s,s'}$ here. 

First, considering $|s_{jk}^+|^2$ gives factor $\delta_{s+} \delta_{s'-}$, the summation about spin fixes $s=+$ and $s'=-$.

Second, the summation about the guiding center $X$ only acts on the part including $\phi_{nX}$. 
Here,
\begin{align}
    \phi_{nX}^\dagger(\vb*{0})\phi_{n^\prime X^\prime}(\vb*{0})&=\frac{1}{2L_y}[ \mathrm{sgn}(nn^\prime)u_{|n|-1}(-X)u_{|n^\prime|-1}(-X^\prime) \non \\
    &\quad +u_{|n|}(-X)u_{|n^\prime|}(-X^\prime) ]. 
\end{align}
The summation of guiding center $X$ be approximated as the integral in the following:
\begin{align}
    \sum_X=\frac{1}{\Delta X}\sum_{X}\Delta X\to\frac{L_y}{2\pi\ell_{\mathrm{B}}^2}\int dX,
\end{align}
So that we obtain:
\begin{align}
    \left(\frac{L_y}{2\pi\ell_{\mathrm{B}}^2}\right)^2
    \int dX\int dX^\prime   |\phi_{nX}^\dagger(\vb*{0})\phi_{n^\prime X^\prime}(\vb*{0})|^2
    =\left(\frac{1}{2\pi\ell_{\mathrm{B}}^2}\right)^2\frac{1}{2}.
\end{align}

\begin{widetext}
    Finally, with the above relation, the $\mathrm{Im}\chi^{R}_{\mathrm{loc}}$ is written as:
\begin{align}
    \mathrm{Im}\chi^{R}_{\mathrm{loc}}(\omega)&=\frac{\pi}{2}\left(\frac{1}{2\pi\ell_{\mathrm{B}}^2}\right)^2\sum_{n,n^\prime}
    \left[f_{\mathrm{FD}}(\varepsilon_{n+})-f_{\mathrm{FD}}(\varepsilon_{n+}+\hbar\omega)\right]\delta(\varepsilon_{n+}-\varepsilon_{n^\prime-}+\hbar\omega) \non \\
    &=\frac{\pi}{2}\left(\frac{1}{2\pi\ell_{\mathrm{B}}^2}\right)^2\sum_{n,n^\prime}
    \int d\varepsilon \left[f_{\mathrm{FD}}(\varepsilon)-f_{\mathrm{FD}}(\varepsilon+\hbar\omega)\right]\delta(\varepsilon-\varepsilon_{n+}) \delta(\varepsilon-\varepsilon_{n^\prime-}+\hbar\omega) \non \\
    &=\frac{\pi}{2}\int d\varepsilon [f_{\mathrm{FD}}(\varepsilon)-f_{\mathrm{FD}}(\varepsilon+\hbar\omega)]
    \left(\frac{1}{2\pi\ell_{\mathrm{B}}^2}\right)^2\sum_{n,n^\prime}\delta(\varepsilon-\varepsilon_{n+})\delta(\varepsilon-\varepsilon_{n^\prime-}+\hbar\omega) \non\\
    &=\frac{\pi}{2}\int d\varepsilon
    [f_{\mathrm{FD}}(\varepsilon)-f_{\mathrm{FD}}(\varepsilon+\hbar\omega)]
    D^{\mathrm{LL}}_+(\varepsilon)
    D^{\mathrm{LL}}_-(\varepsilon+\hbar\omega),
\end{align}
Just as same as Eq.~(15) in the main text.
The density of states with spin index $s$, according to this process, is originally defined as:
\begin{align}
    D^{\mathrm{LL}}_{s}(\varepsilon)=\frac{1}{2\pi\ell_{\mathrm{B}}^2}\sum_n\delta(\varepsilon-\varepsilon_{ns}),
\end{align}
it becomes the form of Eq.~(16) of the main text after considering the level broadening of the Landau Levels and gives a finite value.

\subsubsection{Under Plane Wave Basis}
In the absence of the external magnetic $B$, the electron of the graphene now be described by the plain wave states as eigenstates shown in Eqs.~(\ref{eP}) and (\ref{LP}). Similar to the presence of the magnetic field situation, the summation now decomposed into $\sum_{n,n'}\sum_{\vb*k,\vb*k'}\sum_{s,s'}$, the $|s_{jk}^+|^2$ fixed the $s=+$ and $s'=-$, and the $n(n')$ only takes $\pm$ in this situation. We have:
\begin{align}
    \phi_{n \vb*k}^\dagger(0)\phi_{n^\prime \vb*k^\prime}(0)=\frac{1}{2L_xL_y}\left(1+nn^\prime e^{-i(\varphi_{\vb*k}-\varphi_{\vb*k^\prime})}\right),
\end{align} 
thus, 
\begin{align}
    \sum_{{\vb*k},{\vb*k}^\prime}|\phi_{n{\vb*k}}^\dagger(0)\phi_{n^\prime {\vb*k}^\prime}(0)|^2&=\frac{1}{L_x^2L_y^2}\sum_{{\vb*k},{\vb*k}^\prime}\frac{1}{4}|1+nn^\prime e^{-i(\varphi_{\vb*k}-\varphi_{{\vb*k}^\prime})}|^2 \non \\
    &=\frac{1}{L_x^2L_y^2}\sum_{{\vb*k},{\vb*k}^\prime}\frac{1}{2},
\end{align}
where $\sum_{\vb*k} e^{i\phi_{\vb*k}}=0$ is used.
    With the above relation, we obtain
\begin{align}
    \mathrm{Im}\chi^{R}_{\mathrm{loc}}(\omega)
&=\frac{\pi}{2}\left(
\frac{1}{L_xL_y}
\right)^2\sum_{n,k,n^\prime,k^\prime}
\left[
f(\varepsilon_{n+{\vb*k}})-f(\varepsilon_{n+{\vb*k}}+\hbar\omega)
\right]
\delta(\varepsilon_{n+{\vb*k}}-\varepsilon_{n^\prime-{\vb*k}^\prime}+\hbar\omega) \non \\
&=\frac{\pi}{2}\int d\varepsilon
[f_{\mathrm{FD}}(\varepsilon)-f_{\mathrm{FD}}(\varepsilon+\hbar\omega)]
\left(
\frac{1}{L_xL_y}
\right)^2\sum_{n,{\vb*k},n^\prime,{\vb*k}^\prime}\delta(\varepsilon-\varepsilon_{n+{\vb*k}})
\delta(\varepsilon-\varepsilon_{n^\prime-{\vb*k}^\prime}+\hbar\omega) \non \\
&=\frac{\pi}{2}\int d\varepsilon
[f_{\mathrm{FD}}(\varepsilon)-f_{\mathrm{FD}}(\varepsilon+\hbar\omega)]
D^{\mathrm{PW}}_+(\varepsilon)
D^{\mathrm{PW}}_-(\varepsilon+\hbar\omega),
\end{align}
\end{widetext}
where
\begin{align}
    D^{\mathrm{PW}}_s(\varepsilon) & =\frac{1}{L_xL_y}\sum_{n,{\vb*k}}\delta(\varepsilon-\varepsilon_{n{\vb*k}s}) \non \\
    &=\frac{1}{2\pi(\hbar v)^2}|\varepsilon-J_0Ss|.
\end{align}

\subsection{The Magnon Propagator in FI}
Similar to the dynamic spin susceptibility in graphene $\chi^R$, the retarded Green's function of the magnons in FI is introduced here also as the dynamic spin susceptibility:
\begin{align}\label{GR}
    G^R(\vb*k,t)&:=\frac{1}{i\hbar} \theta(t) \langle[S_{\vb*{k}}^+(t),S_{-\vb*{k}}^-(0)]\rangle_0 \non \\
    &=\frac{1}{i\hbar} \theta(t-t') \langle[S_{\vb*{k}}^+(t),S_{-\vb*{k}}^-(t')]\rangle_0. 
\end{align}
The corresponding Fourier transformation is given by
\begin{align}
    G^R(\vb*k,\omega)&=\int dte^{i(\omega+i0)t}
    \frac{1}{i\hbar}\theta(t)
    \langle[S_{\vb*{k}}^+(t),S_{-\vb*{k}}^-(0)]\rangle_0 \non \\
    &=\frac{2S}{\hbar}\frac{1}{\omega-\omega_{\vb*{k}}+i\alpha\omega},
\end{align}
where the phenomenological damping parameter $\alpha$ is introduced to evaluate the energy levels' broadening. For the magnons, the local spin susceptibility should be the summation of all the components of different wave vectors $\vb*k$:
\begin{align}
    G^R_{\mathrm{loc}}(\omega):=\frac{1}{N}\sum_{\vb*{k}}G^R(\vb*{k},\omega). 
\end{align}
Thus, the imaginary part of the local spin susceptibility is equivalent to the density of states(DOS) of magnons in FI:
\begin{align}
    \mathrm{Im}G^R_{\mathrm{loc}}(\omega)&= \frac{2S}{\hbar} \cdot\frac{1}{N}\sum_{\vb*k} [-\pi \delta(\omega-\omega_{\vb*k})]\non \\
    &=-2\pi S D_m(\hbar \omega),
\end{align}
where the DOS function is defined as
\footnote{
    Here, the following relation is used:
    $\int_0^\infty x^2\delta(X-\alpha x^2)dx=\int_0^\infty \frac{\sqrt{t}}{2\alpha^{3/2}}\delta(X-t)dt=\frac{\sqrt{X}}{2\alpha^{3/2}}
    $}
\begin{align}
    D_m(\varepsilon)&=\frac{1}{N}\sum_k\delta(\varepsilon-\hbar\omega_k) \non \\
    &\to\frac{1}{N}\frac{Na^3}{(2\pi)^3} \int d^3k\delta(\varepsilon-\hbar\omega_k) \non \\
    &=\frac{1}{(2\pi)^3}4\pi\int_0^\infty(ak)^2d(ak)\delta(\varepsilon-\hbar\gamma B-2J_0S(ak)^2) \non \\
    &=\frac{{(2J_0S)^{-3/2}}}{4\pi^2}\sqrt{\varepsilon-\hbar\gamma B},
\end{align}
where the dispersion relation of magnons in FI is just as Eq.~(\ref{DIS}) shown.

\section{Derivation of the Spin Current}
In this appendix, we start from Eq.~(\ref{Is}) to produce the final result composed of the spin susceptibilities above.
The spin current at the interface is generally determined by
\begin{align}
    {I}^{z}_{s} = -\frac{\hbar }{2}\frac{\partial }{\partial t} (s^z_{\mathrm{tot}})=\frac{i}{2} \left[ s^{z}_{\mathrm{tot}}(t),H( t)\right],
\end{align}
where $s^z_{tot}$ is the total spin density just as Eq.~(\ref{Zeemman}) shows
\begin{align}
    s_{\mathrm{tot}}^{z}=\int d\vb*{r} s^{z} (\vb*{r} ).
\end{align}
Thus, the spin current depends on the commutator between $s^{z}(\vb*{r},t)$ and $H_{T}$,
\begin{align}\label{is}
    I_{s}^{z} 
     & =\frac{i}{2}\int d\vb*{r} \left[ s^{z} (\vb*{r}) ,H_{T}\right]  \nonumber\\
     &=- \frac{i}{2}\int d\vb*{r} \sum_{j} J(\vb*{r},\vb*{r}_j)  \left[ s^{+} (\vb*{r})S^-_{j}-\operatorname{H.c.} \right] .
\end{align}
The corresponding statistic average of the interface spin current is
\begin{align}
    \langle I_{s}^{z} \rangle 
    &=2\Im [\int d\vb*{r} \sum_{j} J(\vb*{r},\vb*{r}_j) \langle s^{+} (\vb*{r})S^-_{j}\rangle ],
\end{align} 
 Note that here we add a factor 2 in front of the quantum expectation of Eq.~(\ref{is}) considering both the electrons around $K$ and $K'$ points in graphene give the same contribution (valley degeneracy). Under the interaction picture, considering the first-order and second-order perturbation, we have
\begin{align}
    &\quad \langle s^{+} (\vb*{r} )S_{j}^{-} \rangle \non \\
    &= \langle \psi(t)|s^{+}(\vb*{r} ,t){S}_{j}^{-}( t) |\psi(t)\rangle  \nonumber\\
    & = \langle \psi(0)|{T}_{C}s^{+}(\vb*{r} ,t){S}_{j}^{-}( t) U_{C}|\psi(0)\rangle  \nonumber\\
    & \simeq  \langle  {T}_{C} {s}^{+}(\vb*{r} ,t) {S}_{j}^{-}( t)\left( 1-\frac{i}{\hbar }\int _{C}{H_{T}}( t') dt'\right)\rangle_0  \nonumber  \\
    &= -\frac{i}{\hbar }\int _{C} \langle {T}_{C}[{s}^{+}( \vb*{r} ,t_+){S}_{j}^{-}( t_-) {H_{T}}( t')]\rangle_0 dt',
\end{align}
where the $t_{\pm}$ as shown in Fig.\ref{f1} represent the time points on forward and backward branches $c_{\pm}$ for the same time $t$. 
\begin{widetext}

Substituting ${H}_{T}$ into the spin current expectation
\begin{align}
        \left< I_{s}^{z} (t )\right>  & =2 \operatorname{Im} [\int d\vb*{r}\sum _{j} J(\vb*{r} , \vb*{r}_{j} )\langle s^{+} (\vb*{r} ,t)S_{j}^{-} \rangle ]  \nonumber \\
         & =\frac{1}{\hbar }\int _{C}dt' \int ^{( 2)} d\vb*{r}d\boldsymbol{r'}\sum _{j} J(\vb*{r}, \vb*{r}_{j} )\sum _{j'} J(\vb*{r} ' , \vb*{r}_{j'} ) \operatorname{Re}\left[ \langle {T}_{C}{s}^{+}\left(\vb*{r} ,t_{+}\right){S}_{j}^{-}\left( t_{-}\right){s}^{-} (\vb*{r} ',t'){S}_{j'}^{+}( t') \rangle_0 \right] .
\end{align} 

After applying the Wick expansion for the above equation, it gives
\begin{align}\label{wick}
    \left< I_{s}^{z} (t)\right>  & =\frac{1}{\hbar }\int _{C} dt' \int ^{( 2)} d\vb*{r}d\boldsymbol{r'}\sum _{j} J(\vb*{r} , \vb*{r}_{j} ) \sum _{j'} J(\vb*{r} ', \vb*{r}_{j'}) 
    \operatorname{Re}\left[ \langle {T}_{C}{s}^{+}\left(\vb*{r} ,t_{+}\right)\tilde{s}^{-} (\vb*{r} ',t')\rangle_0 \langle {T}_{C}{S}_{j}^{-}\left( t_{-}\right){S}_{j'}^{+}( t') \rangle_0 \right].
\end{align}
For the $S^{\pm}_{j(j')}$ represents the spin lifting in FI, according to Eqs.~(\ref{sk_1}) and (\ref{sk_2}), are switched to the plane wave basis in the following:
\begin{align}\label{FOR}
    \langle {T}_{C}{S}_{j}^{-}\left( t_{-}\right){S}_{j'}^{+}( t') \rangle_0=\frac{1}{N} \sum_{\vb*{k}} e^{-i(\vb*{r}_{j} -\vb*{r}_{j} ')  \cdot \vb*{k}} \langle {T}_{C}{S}_{\vb*{k}}^{+}(t') {S}_{-\vb*{k}}^{-}\left( t_{-}\right) \rangle_0.
\end{align}
Both the two propagators $\langle {T}_{C}{s}^{+}\left(\vb*{r} ,t_{+}\right){s}^{-} (\vb*{r} ',t')\rangle_0$ and $\langle {T}_{C}{S}_{\vb*{k}}^{+}(t') {S}_{-\vb*{k}}^{-}\left( t_{-}\right) \rangle_0$ could be regarded as following definition Eq.~(\ref{pro})
\begin{align}
    \frac{i}{\hbar}\langle {T}_{C}{s}^{+}\left(\vb*{r} ,t_{+}\right){s}^{-} (\vb*{r} ',t')\rangle_0 &=\chi(\vb*{r},\vb*{r}';t_+,t'),\\
    \frac{1}{i\hbar}\langle {T}_{C}{S}_{\vb*{k}}^{+}(t') {S}_{-\vb*{k}}^{-}\left( t_{-}\right) \rangle_0&=G(\vb*{k};t',t_-).
\end{align}

Substituting all the above representations into Eq.~(\ref{wick}) and use the approximation Eq.~(\ref{approx}), we have:
\begin{align}
    \left< I_{s}^{z} (t)\right>  & =\frac{\hbar J_2^2l^2 A }{N}\sum_{\vb*{k}} \operatorname{Re}\left[ \int _{C} \chi_\mathrm{loc}(t_+,t') G(\vb*{k};t',t_-)  dt'\right] ,
\end{align}
where the $A=L_xL_y$ is the area of the system given by the integral of the free variable $\vb*{r}'$ and $\chi_\mathrm{loc}(t,t')=\chi(\vb*{r}=0;t,t')$.

Further, the integral of the product of the propagators on the Keldysh contour could be rewritten in the following steps to finally reach the formal consists of the spin susceptibilities $\chi^R$ and $G^R$:
\begin{align}
\quad \int _{C} \left[  \chi_\mathrm{loc}(t,t') G(\vb*{k},t) \right] dt' 
    & =\left(\int _{C_+} +\int _{C_-}\right) \chi_\mathrm{loc}(t_+,t') G(\vb*{k};t',t_-) \ dt' \non \\
    &=\int _{-\infty }^{+\infty } [\chi_\mathrm{loc}(t_+,t'_+)G(\vb*{k};t'_+,t_-)-\chi_\mathrm{loc}(t_+,t'_-)G(\vb*{k};t'_-,t_-)] dt',
 \end{align}
For $G(\vb*{k};t'_+,t_-)$, $t'_+$ on the forward branch always appears in front of $t_-$ on the backward branch, thus, the lesser(greater) component $G^{<}(\vb*{k};t,t')=-G^{>}(\vb*{k};t',t)=G(\vb*{k};t'_+,t_-)$ are introduced here. In the same way, when $t'$ and $t$ settle on the same branch, we can decompose the propagator:
\begin{align}
    G(\vb*{k};t'_-,t_-)= \theta(t-t') G^{>}(\vb*{k};t,t')+\theta(t'-t)G^{<}(\vb*{k};t,t'),
\end{align} 
We apply similar operation on $\chi_{\mathrm{loc}}$ and get:
\begin{align}
    \int _{C} \left[  \chi_\mathrm{loc}(t,t') G(\vb*{k};t',t) \right] dt'
    &=\int_{-\infty }^{+\infty } [\chi^{<}_{\mathrm{loc}}(t,t')G^A(\vb*k;t',t)+G^{<}(\vb*{k};t',t)\chi^{R}_{\mathrm{loc}}(t,t')]dt' \non\\
    &=\int_{-\infty }^{+\infty } \frac{d\omega}{2\pi}[\chi^<_{\mathrm{loc}} (\omega)G^{A}(\vb*k,\omega)+\chi^R_{\mathrm{loc}} (\omega)G^{<}(\vb*k,\omega)],
\end{align}
where we use the relationship between the retard (advance) components and greater(lesser) components of the Green's function
\begin{align}\label{green 1}
    G ^{R}({\vb*{k}};t,t') :=\frac{-i}{\hbar } \theta ( t -t')\langle \left[ S^+_{\vb*{k}}( t) ,{S}_{\vb*{k}}(t')\right]\rangle _{0} = \theta ( t -t')[G^{>}({\vb*{k}};t,t') -G ^{< }(\vb*{k};t,t')] ,\\
    G ^{A}({\vb*{k}};t,t') :=\frac{+i}{\hbar } \theta ( t -t')\langle \left[ S^+_{\vb*{k}}( t) ,{S}_{\vb*{k}}(t')\right]\rangle _{0} = \theta ( t -t')[G ^{< }({\vb*{k}};t,t') -G ^{>}({\vb*{k}};t,t')]. 
\end{align}
Then, 
\begin{align}
    \operatorname{Re}\left[ \int _{C} \chi_\mathrm{loc}(t_+,t') G(\vb*{k};t',t_-)  dt'\right] =2 \int _{-\infty }^{+\infty } \frac{d\omega}{2\pi} \operatorname{Im}\chi_{\mathrm{loc}}^R(\omega)[- \operatorname{Im}G^R(\vb*k,\omega) ][f^\mathrm{G}(\omega)-f^\mathrm{FI}(\omega)],
\end{align}
where using the relation gathered from the Lehmann expression of the Green's functions
\begin{align}
    \chi_{\mathrm{loc}}^<(\omega)&=f^{\mathrm{G}}(\omega)\left[2i\mathrm{Im}\chi_\mathrm{loc}^R (\omega)\right] ,\\
    G^<(\vb*k,\omega)&=f^{\mathrm{FI}}(\omega)\left[2i\mathrm{Im}G^R(\vb*k,\omega)\right].
\end{align}
And $[f^\mathrm{G}(\omega)-f^\mathrm{FI}(\omega)]$ represents the distribution functions difference between the two parts of the system. $f^\mathrm{G}(\omega)$ corresponding to the spin excitation in graphene and $f^\mathrm{FI}(\omega)$ corresponding to the magnons in FI. Both of them are Bose-Einstein distribution functions.
If the temperature difference between the two layers of the system (the graphene and the FI) is small enough, then $\frac{\partial f_{\mathrm{BE}}(\omega,T )}{\partial \ T} \delta T$ could replace $[f^\mathrm{G}(\omega)-f^\mathrm{FI}(\omega)]$ where $\delta T=T_{\mathrm{G}} -T_{\mathrm{FI}}$ with
\begin{align}
\frac{\partial f_{\mathrm{BE}}( \omega ,T)}{\partial \ T}
 =-\frac{e^{\hbar \omega /k_{\mathrm{B}}T}}{(e^{\hbar \omega/k_{\mathrm{B}}T}-1)^2}
\left(-\frac{\hbar \omega}{k_{\mathrm{B}}T^2}\right) 
=\frac{k_{\mathrm{B}}}{\hbar \omega}
\frac{(\hbar \omega/2k_{\mathrm{B}}T)^2}{\sinh^2(\hbar \omega/2k_{\mathrm{B}}T)}.
\end{align}
Notice that bosons with no interaction like magnons always have $\mu \equiv 0 $.  

Finally, the spin current is simplified as
\begin{align}
    \left< I_{s}^{z} (t)\right>  
    & ={2\hbar J_2^2l^2 A } \int _{-\infty }^{+\infty } \frac{d\omega}{2\pi} \operatorname{Im}\chi_{\mathrm{loc}}^R(\omega)[- \frac{1}{N}\sum_{\vb*{k}}\operatorname{Im}G^R(\vb*k,\omega) ][f^\mathrm{G}(\omega)-f^\mathrm{FI}(\omega)] \non\\
    &={2 J_2^2l^2 A k_{\mathrm{B}} \delta T }\int _{0 }^{E_M }  d(\hbar \omega) \operatorname{Im}\chi_{\mathrm{loc}}^R(\hbar \omega) [-\operatorname{Im}G^R_{\mathrm{loc}}(\hbar\omega)][\frac{1}{\hbar \omega}
    \frac{(\hbar \omega/2k_{\mathrm{B}}T)^2}{\sinh^2(\hbar \omega/2k_{\mathrm{B}}T)}],
\end{align} 
where the $E_M$ is the artificial cut-off of the assumptive magnon dispersion Eq.~(\ref{DIS}). 

\end{widetext}

\section{The spin pumping spin current}
In the main text, we use the numerical result of the spin current in the same system but generated by the SP effect as a comparison reference to help reveal the mechanism of the spin Seebeck effect. To generate the spin current, the SP way using the irradiation microwave to replace the temperature gradient to motivate the spin current, Just as the previous research\cite{ominato2020quantum,ominato2020valley} has investigated, the spin current at the interface of the graphene/FI bilayer system following the same form of Eq.~(\ref{general}), but with:
\begin{align}\label{SPX}
    \mathrm{Im}\chi^R_{\mathrm{loc}}(\omega) = \frac{\pi}{2}\int d\varepsilon[f_{\mathrm{FD}}(\varepsilon)-f_{\mathrm{FD}}(\varepsilon+\hbar\omega)]
    D^\alpha_+(\varepsilon)D^\alpha_-(\varepsilon),
\end{align}
where $[f_{\mathrm{FD}}(\varepsilon)-f_{\mathrm{FD}}(\varepsilon+\hbar\omega)] \approx \hbar \omega( -\frac{\partial f_{\mathrm{FD}} (\varepsilon )}{\partial \varepsilon })$ and following the same definition of density of states $D^\alpha_s$. And as for the $\delta f_{\mathrm{BE}}=[f^{\mathrm{G}}(\omega)-f^{\mathrm{FI}}(\omega)]$ now becomes:
\begin{align}
    \delta f_{{\boldsymbol{k}}}^{\mathrm{FI}} (\omega )
    &=\frac{\delta G_{{\boldsymbol{k}}}^{< } (\omega )}{2i\operatorname{Im} G_{{\boldsymbol{k}}}^{R} (\omega )} \non \\
     & =\frac{2\pi SN( \gamma h_{\mathrm{ac}} /2)^{2}}{\alpha  \omega } \delta _{{\boldsymbol{k}} ,\boldsymbol{0}} \delta (\omega -\Omega).
\end{align}
The above factor $\delta_{\vb*{k},0}$ limites only the $\vb*{k}=0$ mode $\mathrm{Im}G^R_{\vb*{k}=0}(\omega)$ of $\mathrm{Im}G^R_{\mathrm{loc}}(\omega)$ survives and gives:
\begin{align}
    -\mathrm{Im}G^R_{\vb*{k}=0}(\omega)=\frac{2S}{\hbar}\frac{\alpha \omega }{( \omega -\omega _{{\vb*{k}=0}})^{2} +\alpha ^{2} \omega ^{2}}.
\end{align}
In the above, the electron gyromagnetic ratio $\gamma (<0)$ and the amplitude of the microwave radiation $h_{ac}$ are constants that appear in the FI part of the system Hamiltonian with:
\begin{align}
H_{\mathrm{FI}} =\sum _{\boldsymbol{k}} \hbar \omega _{\boldsymbol{k}} b_{\boldsymbol{k}}^{\dagger } b_{\boldsymbol{k}} -h_{\mathrm{ac}}^{+} (t)b_{{\boldsymbol{k}} =\boldsymbol{0}}^{\dagger } -h_{\mathrm{ac}}^{-} (t)b_{\boldsymbol{k}=\boldsymbol{0}},
\end{align}
with $h_{\mathrm{ac}}^{\pm } (t)=\frac{\hbar \gamma h_{\mathrm{ac}}}{2}\sqrt{2SN} e^{\mp i\Omega t}$. 
The Gilbert damping $\alpha$ is the phenomenological dimensionless damping parameter and here we ignore the higher order shift $\delta \alpha$ proportional to the density of states and spin current above shown in Ref.~\cite{ominato2020quantum}.

One can observe that in Eq.~(\ref{SPX}), the dispersion effect is omitted in the second term of the density of states function while the $D^\alpha_-$ term will have $\varepsilon+ \hbar \omega$ as the variable in SSE situation. This omission occurs in the SP because the frequency $\omega$ is forcibly matched to the frequency $\Omega$ of the microwave irradiation to the FI, and such $\hbar \Omega$ is significantly smaller than the energy gap of the Landau levels, leading to its neglect. 

Similar to the previously discussed spin Seebeck effect, we introduce the dimensionless spin current generated by SP
\begin{align}
    I_{n}^{\mathrm{SP}}=\frac{\langle \hat{I}_{S} \rangle _{\mathrm{SP}} }{I_0^{\mathrm{SP}}},
\end{align}
where the normalization factor is defined by
\begin{align}
    I_0^{\mathrm{SP}}
    =\left[\frac{\Omega ( \gamma h_{\mathrm{ac}} /2)^2}{( \Omega-\omega _{0})^{2} +\alpha^{2} \Omega^{2}}\right]
    \frac{8\times 10^{-9} \pi {\hbar S^2 J_2^2l^2A}}{\sqrt{\mathrm{meV}}\mathrm{nm}^4}.
\end{align}

\section{Peak positions of the spin current by spin pumping}\label{PeakPositionsSP}
As shown in the previous paper\cite{ominato2020quantum}, the peak positions of the spin current generated by SP are independent of temperature because only zero-frequency magnons are contributed. 
The peaks appear at crossing points of spin-up and spin-down Landau levels and are determined 
by solving $\varepsilon_{n+}=\varepsilon_{n'-}$, leading to
\begin{align}
    \frac {1}{B}=\frac{2e\hbar v^2}{(2J_0S)^2}[\operatorname{sgn}(n)\sqrt{|n|}-\operatorname{sgn}(n')\sqrt{|n^{\prime}|}]^2.
\end{align}
Thus, the period of the peak positions is given by 
\begin{align}
\Delta\left(\frac1B\right)=\frac{2e\hbar v^2}{\left(2J_0S\right)^2}.    
\end{align}

\bibliography{ref_rev}

\end{document}